\documentclass[11pt,journal]{article}
\usepackage{cite}
\usepackage{amsfonts}

\usepackage{graphicx}
\usepackage{subfig}

\usepackage{amsmath}

\begin{document}
%
\title{Beam Detection Based on\\Machine Learning Algorithms}
%
%

\author{Haoyuan Li, {hyli16@stanofrd.edu}
        Qing Yin, {qingyin@stanford.edu}}

\maketitle

\begin{abstract}
The positions of free electron laser beams on screens are precisely determined by a sequence of machine learning models. Transfer training is conducted in a self-constructed convolutional neural network based on VGG16 model. Output of intermediate layers are passed as features to a support vector regression model.
With this sequence, 85.8\% correct prediction is achieved on test data.
\end{abstract}


Beam Detection, SVM, CNN

\section{Introduction}

The free electron laser(FEL) at Stanford Linear Accelerator Center(SLAC) is an ultra-fast X-ray laser. As one of the most advanced X-ray light source \cite{felnotes} \cite{RevModPhys.88.015006}, it is famous for its high brightness and short pulse duration: it is 10 billion times brighter than the world's second brightest light source; the pulse duration is several tens femtoseconds.It plays a pivotal role in both fundamental science research and applied research\cite{RevModPhys.88.015006}. The mechanism behind this laser is very delicate\cite{felnotes}. Thus to keep the laser in optimal working condition is challenging.The positions of the electron beams and the laser beams are of fundamental importance in the control and maintenance of this FEL.\\ 

\begin{figure}[!t]
\centering
\includegraphics[width=3in]{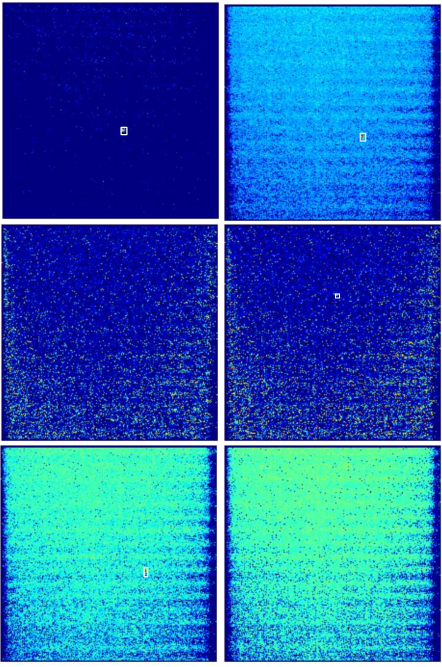}
\caption{Raw photos}
\label{origin}
\end{figure}

Currently, the task of locating beam spots heavily depends on human labor. This is mainly attributed to the wide varieties of beam spots and the presentation of strong noises as demonstrated in Figure~\ref{origin}, where the white square marks the boundary of the beam spot. Each picture requires a long sequence of signal processing methods to mark the beam position. To make things even worse, different instrument configurations and working conditions require different processing parameters. Within the frame of same configuration, the parameters will also drift away along with time advancement because of the inherent delicacy of the instrument. This makes tuning and maintaining the FEL a tedious and burdensome task for researchers.Currently, the data update frequency of the laser is $120\ Hz$. We can barely handle this. In 2020, after the scheduled update, the frequency will climb to $5000\ Hz$.   Thus the only hope lies in automatic processing methods. Considering that simple signal processing method can not handle such nasty condition, We hope to come up with a general machine learning algorithm which is capable of locating the beams' positions quickly and automatically. In this report, we demonstrate  the consecutive application of neural network and supportive vector machine (SVM) to determine the positions of beam spots on  virtual cathode camera (VCC) screen in simple cases.

\section{Challenges and Strategies}
A bird view of such a simple-looking task reveals the challenges behind it. \\
\begin{itemize}
\item Complicated background noises. The background noises are not static. In contrast, it is coherent in both time and space domain due to quantum effect. Thus we have to dynamically change the background when doing background subtraction for different images.
\item Large variety in the intensity and shape of the beam spot. As is indicated above, the maximum intensity of the beam is incredibly high. However, it can also be $0$ which corresponds to the case that no beam spot presented. The shape of the beam can also varies significantly from case to case.

\item Ground truths is hard to find. The signal processing methods fail in finding the beam spot sometimes. There are also cases when the photo is so terrible that even human can not find the beam spot after all the signal processing procedures.
\end{itemize}

The task is similar to face recognition and the expectation of solving it in one strike is unrealistic. Thus, we develop the following strategy. First, we develop an algorithm that is capable of dealing with cases where we have implemented signal processing to realize the fundamental functions. Second, extend the application to cases that we gradually ignore pre-processing steps, e.g. Gaussian blur. In the end, extend the application to raw images. In the essay, we report the accomplishment of the first step that realizes fundamental functions.

\section{Related Work}
A similar dilemma stated above occurred when researchers tried to pinpoint the electron bunch's trace\cite{SLACweb}. It was solved by first feeding the images into a convolutional neural network\cite{CNNweb}, VGG16\cite{Simonyan2014}, and then performing linear regression on its codewords. This gives us the inspiration of this project. Signal processing algorithms have been developed for targeting the beam spot. Thus, we use the results from these algorithms as the ground truth for our training.

\section{Dataset}

Training dataset for this project consists of 162 original VCC screen figures and 16200 figures generated from the original ones (each original image corresponding to 100 generated images). Each figure contains only one beam spot. To generate new figures, we first cut out part of figure that contains the beam spot, then rotate the small patch of figure to some random degree, and put the patch on a random point. In the end, cut out the covered part of the figure and paste it to the original position of the beam spot. The position of the beam on each of the original figures can be determined by the signal processing algorithm. The position of the beam on the generated figures are determined by the generation process.The position of beam in each figure is represented by the coordinates of two diagonal vertices: $z^{(i)} =(y_{\text{min}}^{(i)},y_{\text{max}}^{(i)},x_{\text{min}}^{(i)},x_{\text{max}}^{(i)})^T$

\section{Method}
\subsection{Pipeline}
To obtain an algorithm robust to variations of beam spot and noises, we utilize a two step method. \\

\begin{figure}[!t]
\centering
\includegraphics[width=3in]{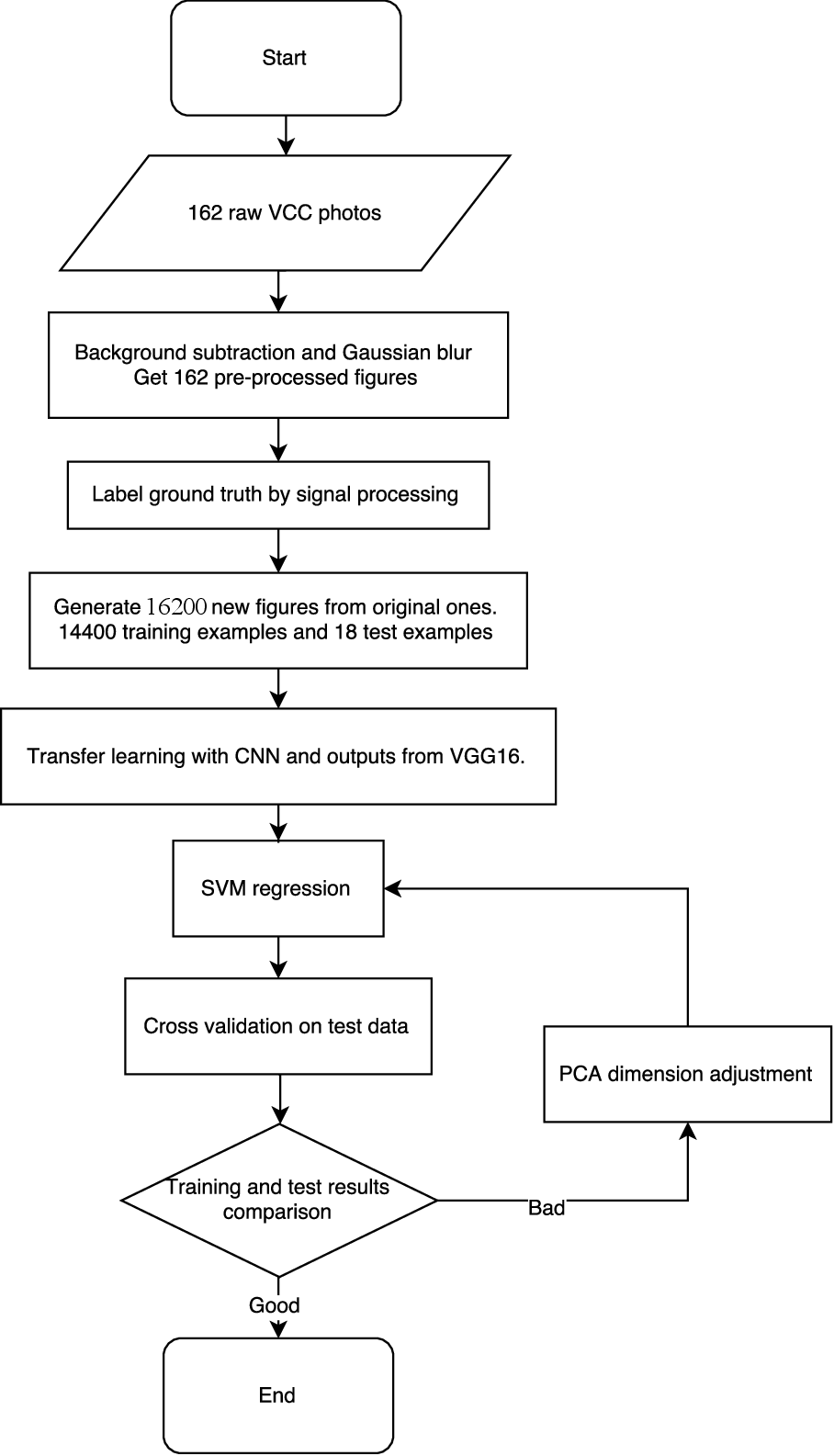}
\caption{Algorithmic flowchart}
\label{pipeline}
\end{figure}

First, we use a convolutional neural network (CNN) as a feature extractor and we take the intermediate outputs of the CNN as features and train a Supportive Vector Regression(SVR) program with them. The pipeline is demonstrated in Figure ~\ref{pipeline}.
\subsection{Feature Extraction}
Since we have moderate sample size, transfer training seems suitable for this purpose. VGG16 is selected as the pre-trained model.\\

\begin{figure}[!t]
\centering
\includegraphics[width=3in]{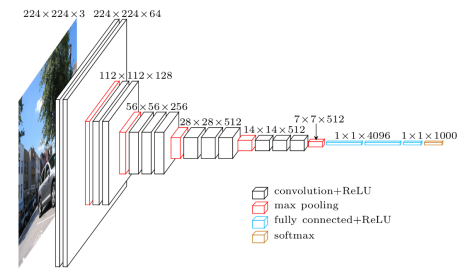}
\caption{Structure of VGG 16\cite{VGGweb}}
\label{cnn}
\end{figure}

VGG16 is a deep convolutional neural network. There are totally 13 layers of convolutional layer, 5 pooling layer and 3 full connected layer. The structure of this network in demonstrated in Figure ~\ref{cnn}.\\

The output of the first and second fully connected layers are both $4096$-dimensional vectors. We feed in our samples and use these two vectors as the features of the image. \\

To do transfer training, first we re-construct the VGG16 with $python\ 2.7$ and $tensorflow\ 0.11$. The last layer of the original VGG16 is a classifier of a thousand classes. We modify it to output the position of the center of the beam spot. The loss is defined as the squared distance between the predicted beam center and the true beam center. A $l^2$ regularization term is added which contains parameters of the last three fully connected layer. 
\begin{equation}
\mathcal{L}_{CNN,loss}=\frac{1}{m}\sum_{i=1}^m||\vec{x}_ii-\vec{\tilde{x}}_i||^2+\frac{\lambda}{m}||W||^2
\end{equation}
where $\vec{x}_i$ and $\vec{\tilde{x}}_i$ indicate respectively the predicted center of the beam and the true center of the beam which is represented by the center of the box. $W$ is a vector consisting of all parameters in the last three full connected layers, i.e. we only tune the parameters in the last three layers. \\

The reason for only tuning the last three layers is that we don't have enough data to tune all parameters. The first several layers, according to modern viewpoint, mainly detect the detailed structures while the upper layers concern more about larger scale objects. So it seems more reasonable to try to tune those layers affecting our next step directly.
 
\subsection{Support Vector Regression}
The feature for the \(i\)th figure is denoted by \(q^{(i)}\in \mathbb{R}^{8192}\).To avoid overfitting, we use $l^2$ regularization when implementing supportive vector regression (SVR) with Gaussian kernel to model the situation. The four parameters of the position are independent. So we train a SVR model for each of the four labels by maximizing the following dual function\cite{smola2004tutorial} 
\begin{equation}
\mathcal{L}_s=\sum_{i=1}^m\alpha_i^sz_s^{(i)}-\frac{1}{2}\sum_{i,j=1}^m \alpha_i^s\alpha_j^s K(q^{(i)},q^{(j)})
\end{equation}
$\alpha_i \in [-t,t], i\in\{1,2,\cdots,m\}$, $s=1,2,3,4$. m is the number of training samples. $\alpha_i^s$'s are the optimization parameters. $t$ is the penalty ratio of the difference between the predicted label and the real label. This dual function is obtained from the primary optimization problem\cite{smola2004tutorial} in the following equations by eliminating the original variables: 
\begin{eqnarray}
\mathcal{L}_s = \frac{1}{2}\lvert\theta_s\rvert_2^2+t\sum_{i=1}^{m}{\xi_i^+}+ t\sum_{i=1}^{m}{\xi_i^-}
\end{eqnarray}
under the constraints that \(\forall i\in\{1,2,\cdots,m\}\)
\begin{eqnarray}
\xi_i^+\geq 0
\end{eqnarray}
\begin{eqnarray}
\xi_i^-\geq 0
\end{eqnarray}
\begin{eqnarray}
z_s^{(i)}- \theta^Tq^{(i)}_{s}\leq \xi_i^+
\end{eqnarray}
\begin{eqnarray}
\theta^Tq^{(i)}_{s}-z_s^{(i)}\leq \xi_i^-
\end{eqnarray}
where \(\xi_i^+\) and \(\xi_i^-\) are the threshold for training error.
\section{Training}
\subsection{Convolutional Neural Network}
Nesterov momentum method\cite{Bengio2012} is utilized to minimize the CNN loss function. We use a vanishing learning rate which decays as $t^{-1}$:
\begin{equation}
\alpha =10^{-4}\times(1+t)^{-1}
\end{equation}
\subsection{Cross Validation}
To maximize the usage of limited data, we use 9-fold cross-validation. First we randomly shuffle the 162 original images to erase the potential correlation between different images. Then divided them into 9 batches, each of 18 images.
For each batch, the other 144 original images together with the related 14400 generated images form the training data set for the next two steps. This guarantees the independence between the training and testing process.

\subsection{PCA and SVM}
 Sequential minimal optimization (SMO) algorithm is utilized to minimize the SVR loss function. Package \emph{scikit-learn} is used throughout the process to guarantee the performance. \\ 
 
The feature has a huge dimension of $8192$, while the prediction only contains $4$ numbers. To prevent over-fitting, we use principle component analysis (PCA) to extract the most influence dimensions and iterate the training and testing procedure for 12 different different dimensions equally distributed in \emph{log}-scale, i.e. $\{5,10,20,40,80,160,320,640,1280,2560,5120,8000\}$.
\section{Results and Discussions}
\subsection{Performance}
After re-tuning, the regression algorithm gives superior performance. Beam positions are corrected predicted in 139 of the 162 original images. The following diagram demonstrates the typical performance of the regression on the test set. When the regression algorithm catches the spot, it gives really precise prediction, which implies that the algorithm does have learned how to pinpoint the position of the spot.
\begin{figure}[!t]
\centering

\includegraphics[width=3in]{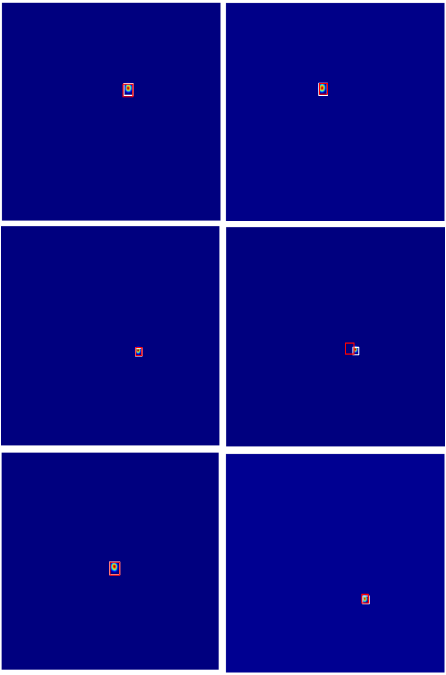}
\caption{Test Results}
\label{test_result}
\end{figure}\\

Yet, when it can not catch the spot (picture in second column of Figure~\ref{test_result}), the regression algorithm just randomly guess a point in the center of the screen. After checking the pictures where the algorithm miss the spot, we found that the missing spots are not dim at all. In fact the maximum point in those figures are even larger than those with perfection predictions. But the beam spots are all very small in those figures.

\subsection{Effect of PCA dimension}
\begin{figure}
\includegraphics[width=3in]{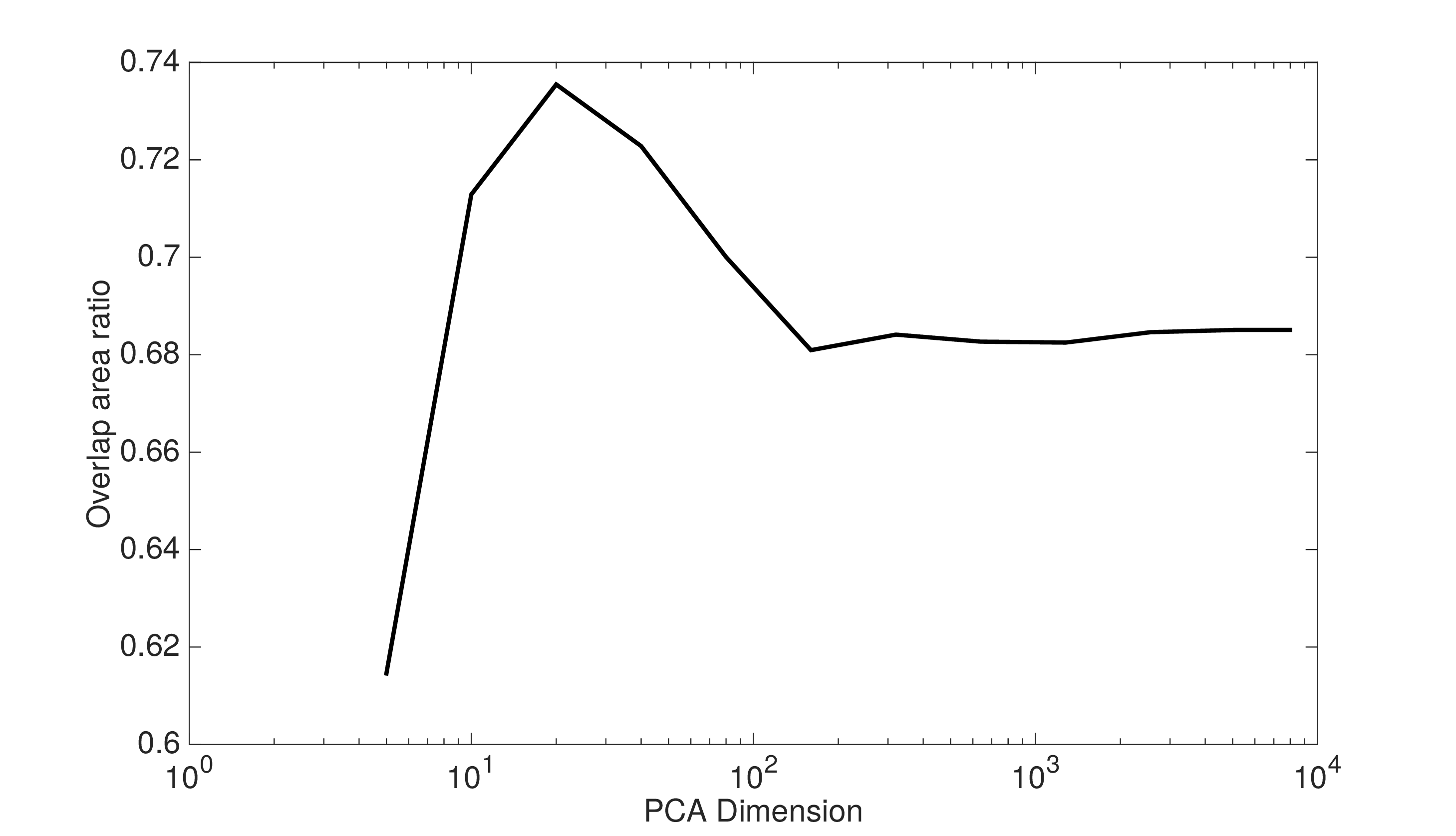}
\caption{Test performance for different PCA dimensions}
\label{PCA_test}
\end{figure}

\begin{figure}
\includegraphics[width=3in]{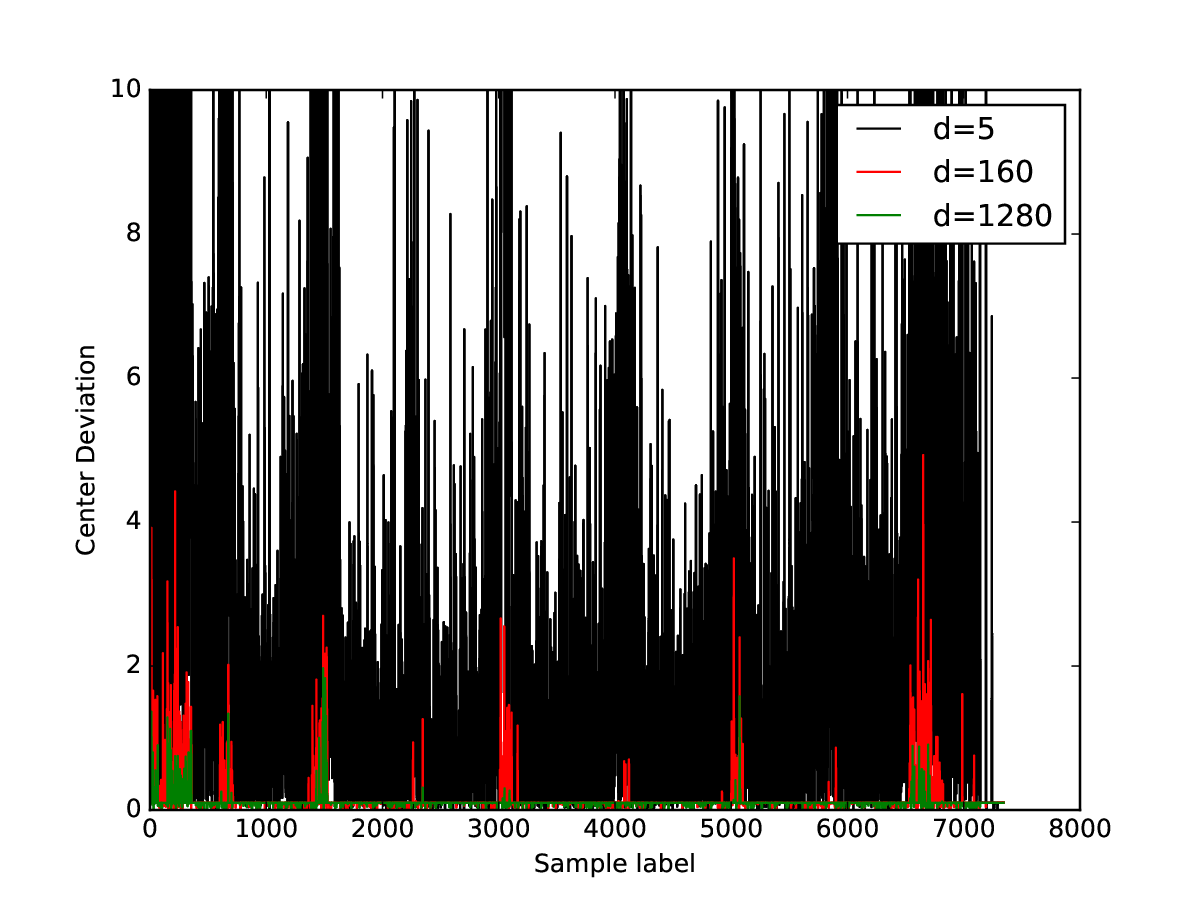}
\caption{Training error for different PCA dimensions}
\label{PCA_training}
\end{figure}

To prevent overfitting, we use PCA to pick out the most influential dimensions of the features before regression. Figure~\ref{PCA_test} demonstrates the training error of the SVR for different PCA dimensions. The error is defined as the distance between the predicted beam center and the true beam center. Figure~\ref{PCA_training} demonstrate the performance of the SVR for different PCA dimensions.To quantify the performance, we use the average ratio between the overlapped area and the area of the true spot as the indicator of the performance.

From these two figures, it is clearly shown that when the PCA dimension exceeds $1000$, there will be severe overfitting problem, as the training vanishes and the performance drops in that area.

\section{Conclusions and Future Works}
With newly generated figures and re-tuned neural network, the performance of the SVR gains great improvement. Yet it is still far from applicable in real environment. Three approaches are scheduled for further improvements.
\subsection{Fine-Tuning Neural Network}
VGG16-like neural networks are huge and hard to tune. It is highly likely that with improved tuning skill and a bit of luck and patience, we can get better performance.
\subsection{More and Harder Data}
\begin{figure}[!t]
\centering
\includegraphics[width=3in]{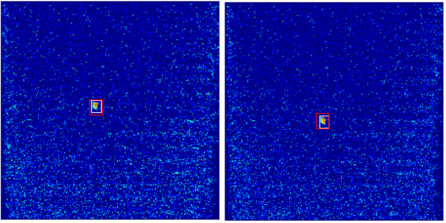}
\caption{Test results for raw pictures}
\label{raw_pic}
\end{figure}

Noisy figures are currently beyond the signal processing algorithm's capability. But as we can see sometimes our regression algorithm has done better job on the existing samples. And even for those raw pictures, our model can generate good prediction results occasionally (Figure~\ref{raw_pic}). So we may try the algorithm on harder samples in the future. Especially we will try to mark the position of the beam by hand for noisy figures and try to train the model on those samples as well.
\subsection{Building New Neural Network}
It is possible that a smaller but better training neural network can do better job. Only after trying both methods can we decide which way is better. We have already constructed another smaller neural network. Training of this one will soon begin. 

\bibliographystyle{abbrv}
\bibliography{main.bib}

\begin{thebibliography}{1}

\bibitem{CNNweb}
{CS231n Convolutional Neural Networks for Visual Recognition}.

\bibitem{SLACweb}
{Spatial Localization - AI at SLAC - SLAC Confluence}.

\bibitem{VGGweb}
{VGG in TensorFlow {\textperiodcentered} Davi Frossard}.

\bibitem{Bengio2012}
Y.~Bengio, N.~Boulanger-Lewandowski, and R.~Pascanu.
\newblock {Advances in Optimizing Recurrent Networks}.
\newblock dec 2012.

\bibitem{felnotes}
K.-J. Kim, Z.~Huang, and R.~Lindberg.
\newblock {\em Synchrotron Radiation and Free-Electron Lasers}.
\newblock Cambridge University Press, 2017.

\bibitem{RevModPhys.88.015006}
C.~Pellegrini, A.~Marinelli, and S.~Reiche.
\newblock The physics of x-ray free-electron lasers.
\newblock {\em Rev. Mod. Phys.}, 88:015006, Mar 2016.

\bibitem{Simonyan2014}
K.~Simonyan and A.~Zisserman.
\newblock {Very Deep Convolutional Networks for Large-Scale Image Recognition}.
\newblock sep 2014.

\bibitem{smola2004tutorial}
A.~J. Smola and B.~Sch{\"o}lkopf.
\newblock A tutorial on support vector regression.
\newblock {\em Statistics and computing}, 14(3):199--222, 2004.

\end{thebibliography}

\end{document}